\documentclass[12pt,preprint]{aastex}
\usepackage{emulateapj5}

\received{}
\accepted{}

\slugcomment{23 January 2002 Revision; Accepted by ApJ Letters}

\begin{document}

\title{FIRST~0747+2739: A FIRST/2MASS Quasar with an Overabundance of
\ion{C}{4} Absorption Systems}

\author{Gordon T. Richards\altaffilmark{1},
Michael D. Gregg\altaffilmark{2},
Robert H. Becker\altaffilmark{2},
and Richard L. White\altaffilmark{3}
}

\altaffiltext{1}{Department of Astronomy and Astrophysics, The Pennsylvania
State University, University Park, PA 16802; gtr@astro.psu.edu}
\altaffiltext{2}{UC-Davis and IGPP/LLNL, L-413, 7000 East Ave., Livermore, CA 94550; gregg,bob@igpp.ucllnl.org}
\altaffiltext{3}{Space Telescope Science Institute, 3700 San Martin Drive, Baltimore, MD 21218; rlw@stsci.edu}

\begin{abstract}

We present a Keck ESI spectrum of FIRST~074711.2+273904, a $K=15.4$
quasar with redshift 4.11 that is detected by both FIRST and 2MASS.
The spectrum contains at least 14 independent \ion{C}{4} absorption
systems longward of the Ly$\alpha$ forest.  These systems are found
over a path length of $\Delta z = 0.984$, constituting one of the
highest densities per unit redshift of \ion{C}{4} absorption ever
observed.  One of the \ion{C}{4} systems is trough-like and resembles
a weak BAL-type outflow.  Two of the \ion{C}{4} are ``associated''
absorption systems with $|v| < 3000\,{\rm km\,s^{-1}}$.  Of the 11
remaining systems with $v > 3000\,{\rm km\,s^{-1}}$, eight are either
resolved or require multiple discrete systems to fit the line
profiles.  In addition to \ion{C}{4} absorption, there are two
low-ionization \ion{Mg}{2} absorption systems along with two damped
Ly$\alpha$ systems, at least one of which may be a \ion{C}{4} system.
The overdensity of \ion{C}{4} absorption spans a redshift range of
$\Delta z \sim 1$.  Superclusters along the line of sight are unlikely
to cause an overdensity stretching over such a long redshift path,
thus the absorption may be an example of narrow, high-velocity,
intrinsic absorption that originates from the quasar.  We suggest that
this quasar is a member of a transitional class of BAL quasars where
we are just barely seeing the spatial, density, or temporal edge of
the BAL-producing region (or period); the multiple high-velocity
absorption systems may be the remnants (or precursors) of a stronger
BAL outflow.  If correct, then some simpler absorption line complexes
in other quasars may also be due to outflowing rather than intervening
material.

\end{abstract}


\section{Introduction}

It is generally agreed that the broad absorption line (BAL) systems
seen in the spectra of many quasars are intrinsic to the quasar,
arising from high velocity outflow of gas directly from the accretion
disk region \citep{wmf+91}.  Narrow absorption lines in quasar
spectra, however, arise from a variety of sources.  Narrow absorption
lines with small velocities relative to the quasar emission redshift
could be caused by galaxies along the line of sight, clouds in the
interstellar medium of the host galaxy, or even smaller scale gas
flows within a few parsecs of the black hole.  On the other hand, most
of the narrow absorption systems that have large velocities with
respect to the emission redshift of the quasar are thought to be
intervening \citep{ssb88}; however, there is increasing evidence that
even some fraction of these may be intrinsic.  In many BAL quasars,
the absorption troughs have considerable structure, suggesting a
connection to narrow absorption line systems.  We discuss the spectrum
of a newly discovered quasar, FIRST J074711.2+273904 (hereafter
FIRST~0747+2739), with an overdensity of \ion{C}{4} absorbers in its
spectrum.  It may be an example of an object whose BAL component has
been caught in a transitional state that could be related to
orientation, time, or density.

\section{Data}

FIRST~0747+2739 was discovered in the early stages of an ongoing
project to search for extremely red quasars \citep{glw+02}.
Candidates are found by matching point sources from the The 2-Micron
All-Sky Survey (2MASS; \citealt{klp+94}) with FIRST radio sources, and
then further matching to an optical catalog such as the APM
\citep{mi92}.  This object also has been discovered independently by
\citet{bvp+02}.  FIRST~0747+2739 has $R-K = 2.5$, not particularly
red, but it is undetected on the blue Palomar sky survey plate,
yielding $B-K \gtrsim 7$.  The initial spectrum was obtained using the
Kast spectrograph on the Lick Observatory 3m telescope.  A higher
signal-to-noise spectrum was obtained at Keck Observatory using the
Low Resolution Imaging Spectrograph (LRIS; \citealt{occ+95}),
revealing a strikingly rich system of absorption lines redward of
Ly$\alpha$.

FIRST~0747+2739 is located at $07^{\rm h} 47^{\rm m} 11^{\rm
s}.208+27^{\circ} 39\arcmin 04\farcs00$ (J2000).  There is a 2MASS
source within $1\arcsec$ with $J$, $H$, and $K$ magnitudes of $16.77$,
$16.16$, and $15.38$, respectively.  The peak $20\,{\rm cm}$ flux
density as measured by FIRST \citep{bwh95} is $1.08\pm0.14\,{\rm
mJy}$; the integrated FIRST flux density is $1.55\,{\rm mJy}$.

Motivated by the LRIS spectrum, we obtained a high-dispersion spectrum
of FIRST~0747+2739 using the echelle spectrograph and imager (ESI;
\citealt{em98}) on the Keck II telescope on the night of 2000 April 6.
The night was photometric with $0\farcs9$ seeing.  A 900s
high-resolution spectrum was taken through a $1\farcs0$ slit in the
echellette mode of ESI.  In this mode, the spectral range of
$3900\,{\rm \AA}$ to $11000\,{\rm \AA}$ is covered in 10 spectral
orders with a nearly constant dispersion of $11.4\,{\rm km}\,{\rm
s^{-1}}\,{\rm pixel^{-1}}$.  The signal-to-noise of the spectrum
ranges from 20 to 40 per pixel.  Wavelength calibrations were
performed with observations of a CuAr lamp.  The spectrophotometric
standard G191-B2B \citep{msb+88,mg90} was observed for flux
calibration.  All observations were carried out at the parallactic
angle.  The data were reduced using a tailored set of IRAF and IDL
routines developed by us specifically for ESI data.

\section{Analysis and Discussion}

\begin{figure}[h]
\epsscale{0.9}
\plotone{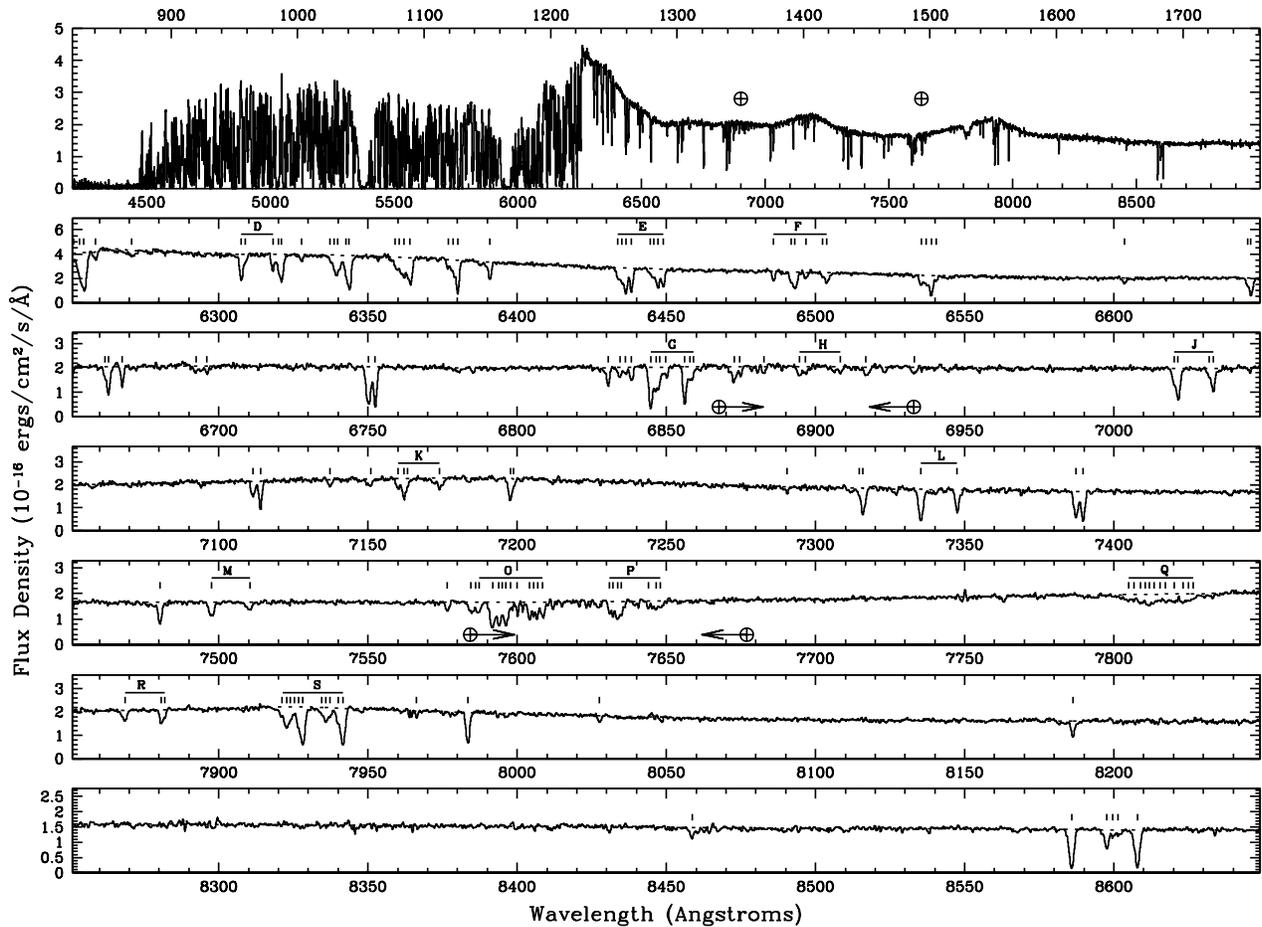}
\caption{({\em Top}) The FIRST~0747+2739 ESI spectrum from
$4000\,{\rm\AA}$ to $9000\,{\rm\AA}$, smoothed by 9 pixels; the top
axis gives the rest frame wavelengths.  Note the weak trough just
blueward of the C\,IV emission line.  Regions that may be affected by
the ${\rm O_2}$ A and B telluric absorption bands are labeled.
However, the absorption in these regions is probably real since the
spectrum of another quasar taken on the same night showed no residual
absorption after telluric correction.  ({\em Bottom Six Panels}) An
expanded view of the spectrum longward of Ly$\alpha$ at the full
$11.4\,{\rm km}\,{\rm s^{-1}}\,{\rm pixel^{-1}}$ resolution.  The 14
C\,IV systems are labeled; a horizontal line connects the wavelength
of the lowest redshift $\lambda\,1548$ to the highest redshift
$\lambda\,1551$ line. \label{fig:fig1}}
\end{figure}

The overdensity of absorption features in the spectrum of
FIRST~0747+2739 is immediately apparent in the Keck ESI data
(Figure~\ref{fig:fig1}).  Also conspicuous are a weak trough-like
feature near $7815\,{\rm \AA}$ and two damped Ly$\alpha$ systems.  The
region longward of Ly$\alpha$ emission is plotted on an expanded scale
in the lower panels of Figure~\ref{fig:fig1}.  We used an automated
routine to search this spectral region for absorption lines with
significance greater than $10\sigma$.  A total of 42 absorption line
systems were identified, 34 being \ion{C}{4} doublets, including a
BAL-like \ion{C}{4} trough.  The systems are given in
Table~\ref{tab:tab1}, which lists the system identification code, the
redshift, the rest equivalent width of the blue and red components of
the \ion{C}{4} doublet (if detected), the ``ejection'' velocity of the
system, and other absorption species belonging to each system.  The
velocities are with respect to the quasar redshift and are given by $v
= (R^2-1)/(R^2+1)$, where $R=(1+z_{em})/(1+z_{abs})$.  The individual
systems are merged into 18 ``Poisson'' systems (14 of which are
\ion{C}{4}) by combining absorbers within $1000\,{\rm km\,s^{-1}}$
\citep{ssb88}.

\subsection{Line Densities}

How unusual is the abundance of \ion{C}{4} absorbers seen towards
FIRST~0747+2739?  At high-redshift, \citet{ste90} found that (based on
11 quasars) the expected number of \ion{C}{4} absorbers per unit
redshift is approximately $1.0\pm0.5$ at $z\sim3.25$ for rest
equivalent width larger than $0.15\,{\rm \AA}$ and $\beta c >
5000\,{\rm km\,s^{-1}}$.  If we sum the equivalent widths of the each
of the \ion{C}{4} systems within $150\,{\rm km\,s^{-1}}$, and ignore
those three systems with $\beta c < 5000\,{\rm km\,s^{-1}}$, we find
seven combined systems with rest equivalent widths larger than
$0.15\,{\rm \AA}$ for both lines in the \ion{C}{4} doublet.  The path
length searched for \ion{C}{4} between Ly$\alpha$ emission and
\ion{C}{4} emission is $\Delta z = 0.984$.  Thus, $dN/dz$ is $7.11$
for this line of sight --- an excess of 12 standard deviations from
the \citet{ste90} measurement.

For weak \ion{C}{4} lines, \citet{tls96} find $dN/dz = 7.1\pm1.7$ for
$W_r > 0.03\,{\rm \AA}$ and $1.5 < z < 2.9$ (based on four quasars).
If we impose constraints on the \citet{tls96} data to match our more
conservative $10\sigma$ significance limit and combine systems within
$200\,{\rm km\,s^{-1}}$, then the density reported by \citet{tls96} is
reduced to $dN/dz = 3.75\pm1.25$.  In our Keck ESI spectrum of
FIRST~0747+2739 we find, using the same criteria, at least 9 distinct
high-velocity \ion{C}{4} systems.  Over the path length observed, that
yields $dN/dz = 9.1\pm3.0$, which is a $4.3\sigma$ excess as compared
to \citet{tls96}; thus there is evidence that the overdensity of
absorption may extend to weaker systems.

Since \citet{ssb88} and \citet{ste90} found that the density of
absorption lines decreases with increasing redshift (for strong lines)
and since the \ion{C}{4} absorption systems found in FIRST~0747+2739
are at higher redshifts than either the \citet{tls96} or \citet{ste90}
samples, our observed density of lines may be even more significant.

What can explain the large overdensity of absorption toward
FIRST~0747+2739?  All possible explanations can be classified as
intervening or intrinsic.  In the intervening case, the absorption
lines arise from material associated with galaxies along the line of
sight that have no relation to the quasar.  In the intrinsic case, the
absorption is caused by high velocity ejecta from the quasar itself.

\subsection{The Intervening Hypothesis}

The number density of strong \ion{C}{4} lines towards this quasar is
much larger than is normally observed.  An examination of the spectra
and tables from \citet{ssb88} and \citet{ste90} along with a query of
the \citet{yyc+91} catalog of quasar absorption lines reveals that
such a concentration of \ion{C}{4} absorption over such a long path
length is extremely rare.  None of the 11 high-redshift quasars from
\citet{ste90} has a density of \ion{C}{4} absorption higher than
FIRST~0747+2739.  Two of the 55 lower redshift quasars from
\citet{ssb88} do have absorption line densities higher then
FIRST~0747+2739; Q0013-004 (UM~224; $z=2.086$) and Q0854+191
($z=1.896$) have six and seven strong ($W_r > 0.15\,{\rm \AA}$)
``Poisson'' systems over path lengths of $\Delta z \sim 0.6$,
respectively.

Quasars that have relatively high densities of absorption have created
much interest in the search for signs of superclustering at high
redshift \citep[e.g.,][]{hhw89,rfs91,fhc+93,di96}.  However, most of
these quasars have absorption systems where the overdensity is very
limited in its redshift range, such as CSO~118 \citep{gcb01} and the
well-studied PKS~0237$-$233 \citep{fhc+93}.  FIRST~074711.2+273904 is
different in the sense that the overdensity of \ion{C}{4} absorption
is spread over a large redshift range ($\Delta z \sim 1$) --- much too
large to be explained by a single supercluster.  A search of the NASA
Extragalactic Database (NED) to look for other quasars to test the
superclustering hypothesis found no other known quasars within 30
arcminutes, although there are five other radio sources within
$10\arcmin$ of FIRST~0747+2739.

One possible way to reconcile the intervening hypothesis with the
absorption line overdensity in FIRST~0747+2739 is if the quasar is
being gravitationally magnified by the intervening material.  Such
magnification could occur without producing multiple images, and might
help to explain why a quasar with $z>4$ is so bright.  The possibility
of compound microlensing of quasars by their absorption line systems
is not a new concept \citep{tw95,vqy+96}.  This object, however, is a
point source in our NIRC K-band image from Keck Observatory, obtained
in 0\farcs4 seeing.

Finally, it is possible that the apparent overdensity of absorption
systems in FIRST~0747+2739 is either just chance coincidence, or is
not actually all that uncommon.  The number of high-redshift quasars
that are well-studied for absorption is still quite small and
additional investigations may yet turn up more cases like
FIRST~0747+2739.  Such a finding would not necessarily be evidence for
either the intrinsic or the intervening hypothesis, however.  For
now, we conclude that the nature of FIRST~0747+2739 is highly unusual
and next to impossible to reconcile with any picture where a
significant number of the absorbing systems can be attributed to
intervening objects unassociated with the quasar.

\subsection{The Intrinsic Hypothesis}

Strong circumstantial evidence to support the intrinsic interpretation
for the plethora of narrow, high-velocity absorption systems in
FIRST~0747+2739 is the presence of similar, almost certainly
intrinsic, absorption at low velocities relative to the quasar
emission line frame.  System S is typical of intrinsic absorption line
systems seen in other quasars, exhibiting \ion{N}{5} absorption and a
small blueshift relative to the quasar's \ion{C}{4} emission redshift.
System Q appears to be a weak BAL-like outflow; System P is also
broad, but the blue and red components of the troughs are resolved.
Finally, System R, with a velocity of only $+1600$ km~s$^{-1}$, may
also be an intrinsic absorption system.

There is ample evidence in the literature for the intrinsic
interpretation of high-velocity non-BAL outflows in quasars.
\citet{jhk+96} and \citet{hbc+97} found weak, high-velocity,
trough-like outflows; \citet{hbc+97} showed that the outflows in
Q~2345+125 are intrinsic because the absorption profiles varied in
$\sim3.5$ months in the quasar rest frame.  \citet{ric01} and
\citet{rlb+01} give statistical evidence that some narrow,
high-velocity \ion{C}{4} absorption line systems may be intrinsic in
nature.  Finally, \citet{gcb01} discuss an overdensity of \ion{C}{4}
absorption in CSO~118.

FIRST~0747+2739 is similar to one of the quasars that comprise the
so-called ``Tololo Pair'' \citep[e.g.,][]{di96}, namely
Tol~1038$-$2712, which also has weak BAL-like troughs.  This
well-studied pair is a favorite for superclustering studies
\citep[e.g.,][]{jpu+86} since the two quasars appear to have common
absorption line systems despite the rather large ($17\arcmin.9$)
separation between the quasars.  Other authors
\citep[e.g.,][]{cds87,rob87} still favor the hypothesis that both are
BALQSOs and the absorption is from intrinsic high velocity outflow.
The velocity separation of the lines in the Tololo Pair is reminiscent
of the of the double-trough BAL structure found by \citet{kvm+93} for
BALQSOs in general, lending credence to the idea that the absorption
lines in the Tololo pair are intrinsic.  In any case, the similarity
of Tol~1038$-$2712 and FIRST~0747+2739 may help lead to an
understanding of the nature of the absorption in both systems.

Q1303+308 provides another interesting comparison; because of its
complex absorption structure, it has been classified as a BALQSO
\citep{fwm+87}; however, the absorption lines are all very narrow and
do not meet the traditional BAL criteria set forth largely by the same
authors a few years later \citep{wmf+91}.  One must then wonder, how
we would classify an object with a slightly lower incidence of
absorption than Q1303+308.  Such an object might simply be classified
as a quasar with an overdensity of absorption line systems much like
FIRST~0747+2739.

The fashionable method of determining whether a particular absorption
system is intrinsic is either to determine the ``coverage fraction''
of the quasar or to show that the absorption line system is variable
\citep{bhs97,hbj+97}.  Lacking the data to test for variability, we
have analyzed the coverage fractions for each of our systems as a
function of velocity.  Many of the high-velocity systems do indeed
have coverage fractions that are consistent with the intrinsic
hypothesis; however, the majority of the systems are resolved into
multiple components or blended, which means that the coverage fraction
results are not entirely reliable.  Thus, on the basis of our covering
fraction analysis alone, we are not able to confirm (or exclude) the
possibility that the overdensity of absorption lines in this quasar
spectrum is due to intrinsic absorption.

Regardless of the lack of conclusive evidence from variability or
coverage fractions, the presence of trough-like absorption at moderate
velocities and the large redshift span of the overdensity of the
\ion{C}{4} absorption points towards an intrinsic origin of at least
some of the relatively narrow, high-velocity \ion{C}{4} absorption
systems.  We suggest that FIRST~0747+2739 is a quasar in transition
from (to) a quasi-BAL object to (from) a standard quasar.  The lack of
strong, broad absorption can be explained as being the result of one
or more effects.  As time goes on, it may be natural for the lines to
become narrow as the clouds dissipate down to their small, dense
cores; or BAL features might be a superposition of many narrow
features and FIRST~0747+2739 has a relatively low density of them; or
perhaps the many narrow absorbers can somehow be attributed to
orientation effects.

\section{Conclusions}

In summary, we find that that overdensity of \ion{C}{4} absorption
lines seen along the line of sight to FIRST~0747+2739 is sufficiently
high that it is likely that at least some of these systems are
high-velocity, intrinsic absorption systems.  The presence of weak
trough-like absorption features suggests that FIRST~0747+2739 may be a
transitional BAL quasar that is being observed at a brief phase in its
transition to (or from) a standard quasar, or is being viewed at a
highly specific orientation, or that it is a BAL with a very low
absorber density.  If the absorption is intrinsic, monitoring with
high resolution spectroscopy over the course of $\gtrsim 18$ months
may reveal variability of the absorbers, a signature of intrinsic
absorption.  Confirmation (or exclusion) of the intrinsic hypothesis
for the unusually large number of high-velocity \ion{C}{4} systems in
FIRST~0747+2739 would have interesting consequences for \ion{C}{4}
absorption line studies, particularly high-redshift superclustering
studies.

\acknowledgements

GTR acknowledges support from NSF grant AST99-00703.  RHB and MDG
acknowledge support from the NSF (grants AST-98-02791 and
AST-98-02732) and the Institute of Geophysics and Planetary Physics
(operated under the auspices of the U.S. Department of Energy by the
University of California Lawrence Livermore National Laboratory under
contract No. W-7405-Eng-48).  This research has made use of the
NASA/IPAC Extragalactic Database (NED) which is operated by the Jet
Propulsion Laboratory, California Institute of Technology, under
contract with the National Aeronautics and Space Administration.  We
thank Rajib Ganguly and Jane Charlton for useful discussions.

\begin{deluxetable}{lrccrl}
\tabletypesize{\scriptsize}
\tablecaption{Absorption Systems\label{tab:tab1}}
\tablewidth{0pt}
\tablecolumns{5}
\tablehead{
\colhead{} & \colhead{} & 
\multicolumn{2}{c}{C\,IV ${\rm W_{r}}$} & 
\colhead{Vel.} & \colhead{} \\
\cline{3-4} \\
\colhead{Sys.} & \colhead{$z$} &
\colhead{$1548$} & \colhead{$1551$} &
\colhead{(km/s)} & \colhead{Other Lines}
}
\startdata
A1a & 1.3762\tablenotemark{a} & & & 193330 & Mg\,II \\
A1b & 1.3766 & & & 193300 & Mg\,II,Mg\,I \\
\vspace{0.05cm}
A2 & 1.3783 & & & 193170 & Mg\,II,Mg\,I \\
B & 2.0704 & & & 140840 & Mg\,II,Mg\,I,Fe\,II 2344, \\
\vspace{0.05cm}
 & & & & & 2374,2383,2587,2600 \\
\vspace{0.05cm}
C & 2.6083\tablenotemark{b} & & & 100370 & Fe\,II 2344,2374,2383 \\
\vspace{0.05cm}
D & 3.0741 & 0.18 & 0.11 & 66830 & \\
E1a & 3.1556\tablenotemark{a} & 0.09 & 0.05 & 61150 & \\
E1b & 3.1565\tablenotemark{a} & 0.14 & 0.08 & 61090 & \\
E1c & 3.1574 & 0.24 & 0.18 & 61030 & \\
\vspace{0.05cm}
E2 & 3.1585 & 0.22 & 0.17 & 60950 & \\
F1 & 3.1893 & 0.09 & 0.06 & 58830 & \\
\vspace{0.05cm}
F2 & 3.1939\tablenotemark{c} & 0.21 & 0.12 & 58510 & \\
G1 & 3.4212 & 0.28 & 0.24 & 43140 & DLA,Si\,II,Fe\,II,Al\,II \\
G2 & 3.4222 & 0.15 & 0.08 & 43070 & DLA \\
\vspace{0.05cm}
G3 & 3.4231 & 0.09 & 0.04 & 43020 & DLA,Si\,II,Fe\,II,Al\,II \\
H1 & 3.4534\tablenotemark{d} & 0.06 & 0.03 & 41000 & \\
\vspace{0.05cm}
H2 & 3.4546\tablenotemark{d} & 0.04 & 0.04 & 40920 & \\
Ja & 3.5344\tablenotemark{a} & 0.10 & 0.05 & 35680 & Si\,IV?,Al\,II? \\
\vspace{0.05cm}
Jb & 3.5353 & 0.21 & 0.14 & 35620 & Si\,IV?,Al\,II? \\
K1 & 3.6248\tablenotemark{d} & 0.06 & 0.03 & 29830 & \\
K2a & 3.6260 & 0.11 & 0.06 & 29750 & \\
\vspace{0.05cm}
K2b & 3.6268\tablenotemark{a} & 0.04 & 0.03 & 29700 & \\
\vspace{0.05cm}
L & 3.7380 & 0.24 & 0.15 & 22630 & \\
\vspace{0.05cm}
M & 3.8428 & 0.11 & 0.05 & 16100 & \\
N1 & 3.8972\tablenotemark{a} & & & 12750 & DLA,O\,I,Si\,II\tablenotemark{e}, C\,II, \\
& & & & & \ion{Fe}{2},\ion{Al}{2} \\
N2 & 3.8985\tablenotemark{a} &  & & 12670 & DLA,O\,I,Si\,II\tablenotemark{e}, C\,II, \\
 & &  & & & \ion{Fe}{2},\ion{Al}{2} \\
N3 & 3.8996 & ? & ? & 12610 & DLA,O\,I,Si\,II\tablenotemark{e}, C\,II, \\
\vspace{0.05cm}
 & & & & & Fe\,II,Al\,II,Si\,IV? \\
O1 & 3.9008\tablenotemark{f} & 0.06 & 0.07 & 12530 & Si\,IV\\
O2 & 3.9036\tablenotemark{f} & 0.20 & 0.11 & 12360 & Si\,IV\\
O3 & 3.9050\tablenotemark{f} & 0.14 & 0.10 & 12280 & Si\,IV\\
\vspace{0.05cm}
O4 & 3.9065\tablenotemark{f} & 0.15 & 0.11 & 12190 & Si\,IV\\
P1 & 3.9289\tablenotemark{g} & 0.06 & 0.04 & 10820 & \\
P2 & 3.9297\tablenotemark{g} & 0.08 & 0.03 & 10770 & \\
P3 & 3.9306\tablenotemark{g} & 0.11 & 0.05 & 10720 & \\
\vspace{0.05cm}
P4 & 3.9315\tablenotemark{g} & 0.08 & 0.04 & 10660 & \\
\vspace{0.05cm}
Q & 4.0449\tablenotemark{h} & BAL? & BAL? & 3850 & \\
\vspace{0.05cm}
R & 4.0825 & 0.06 & 0.03 & 1620 & \\
S1a & 4.1164\tablenotemark{a} & 0.05 & 0.04 & -380 & N\,V \\
S1b & 4.1174 & 0.11 & 0.07 & -430 & N\,V\\
S1c & 4.1182\tablenotemark{a} & 0.07 & 0.05 & -480 & N\,V\\
S2a & 4.1200\tablenotemark{a} & 0.11 & 0.09 & -590 & N\,V\\
S2b & 4.1208 & 0.23 & 0.22 & -630 & N\,V\\
\enddata

\tablecomments{Systems are combined within $1000\,{\rm km\,s^{-1}}$.
The letter ``I'' is skipped to avoid confusion.  Major subsystems
(distinct minima) are indicated by 1, 2, 3, 4 after the system letter.
Minor subsystems (needed to explain the line profile) are indicated by
a, b, c.  Unless otherwise indicated, the absorption line systems are
damped Ly$\alpha$ (DLA); Mg\,II $\lambda\lambda\,2796,2804$; Mg\,I
$\lambda\,2853$; Al\,II $\lambda\,1671$; O\,I $\lambda\,1602$; C\,II
$\lambda\,1335$; Fe\,II $\lambda\,1608$; Si\,II $\lambda\,1527$;
Si\,IV $\lambda\lambda\,1393,1402$; and N\,V
$\lambda\lambda\,1238,1242$.}

\tablenotetext{a}{Weaker system.  Needed to explain profile of
stronger system.}
\tablenotetext{b}{Based on Fe\,II.  Mg\,II is beyond the wavelength range studied.}
\tablenotetext{c}{Line profile is resolved at this resolution.}
\tablenotetext{d}{Questionable system.}
\tablenotetext{e}{Also Si\,II $\lambda\,1304$.}
\tablenotetext{f}{Heavily blended.}
\tablenotetext{g}{Heavily blended, trough-like.}
\tablenotetext{h}{Too blended and trough-like to resolve into components.}
\end{deluxetable}

\end{document}